\documentclass[cameraready]{Interspeech}
\usepackage[T1]{fontenc}

\usepackage{zref-clever}
\zcsetup{S}

\usepackage[labelformat=simple]{subcaption}

\usepackage{mathtools}
\usepackage{amsfonts}
\usepackage{bbm}
\usepackage{cite}
\usepackage{nicematrix}
\usepackage[x11names,table]{xcolor}
\definecolor{lightgreen}{rgb}{0.9, 1.0, 0.9}

\newcommand{\parag}[1]{\noindent\textbf{#1.}}

\newcommand{\styVec}[1]{\mathbf{#1}}
\newcommand{\stySet}[1]{\mathcal{#1}}
\newcommand{\styDec}[1]{\textsc{#1}}
\newcommand{\styModel}[1]{\textsc{#1}}
\DeclareMathOperator*{\argmax}{argmax}

\DeclareMathOperator{\expect}{\mathbb{E}}
\newcommand{\R} {\mathbb{R}}

\newcommand{\Z} {\mathbb{Z}}

\newcommand{\vcbTok}  {\stySet{V}}

\newcommand{\vcbAlph} {\Sigma}
\newcommand{\spaceIn} {\stySet{X}}
\newcommand{\spaceOut}{\stySet{Y}}
\newcommand{\setHyp}{\stySet{H}}
\newcommand{\setPref}{\stySet{R}}
\newcommand{\setSampleCtc}{\stySet{Z}}

\newcommand{\funcUtility} {u}
\newcommand{\funcCtcPath}{\mathcal{B}}
\newcommand{\funcDetok}    {\sigma}
\newcommand{\distBernoulli}[1]{\mathrm{Bernoulli}\left({#1}\right)}
\newcommand{\distCategorical}[1]{\mathrm{Cat}\left({#1}\right)}
\newcommand{\funcWer}{\mathrm{WER}}
\newcommand{\tokIn}  {x}
\newcommand{\tokOut} {y}

\newcommand{\tokCtc} {z}
\newcommand{\tokHyp} {h}
\newcommand{\tokMask} {m}

\newcommand{\seqIn}  {\styVec{\tokIn}}
\newcommand{\seqOut} {\styVec{\tokOut}}

\newcommand{\seqCtc} {\styVec{\tokCtc}}
\newcommand{\seqHyp} {\styVec{\tokHyp}}
\newcommand{\seqMask} {\styVec{\tokMask}}

\newcommand{\blank}{\varnothing}

\newcommand{\thresMask}{\alpha}
\newcommand{\niter}{N_\text{iter}}

\newcommand{\curTime}{t}

\newcommand{\param}{\theta}
\newcommand{\paramCtc}{\param_{\styModel{ctc}}}
\newcommand{\paramCmlm}{\param_{\styModel{cmlm}}}

\newcommand{\decMap}{\styDec{map}}
\newcommand{\decMbr}{\styDec{mbr}}
\newcommand{\decAr}{\styDec{ar}}
\newcommand{\decNar}{\styDec{nar}}

\newcommand{\modelCtc}{\styModel{ctc}}
\newcommand{\modelCmlm}{\styModel{cmlm}}

\newcommand{\styTool}[1]{\texttt{#1}}
\newcommand{\toolEspnet}{\styTool{ESPNet}}
\newcommand{\toolMbrs}{\styTool{mbrs}}

\newcommand{\regmark}{${}^{\text{\textregistered}}$}
\newcommand{\trademark}{${}^{\text{\texttrademark}}$}

\title{
Non-Autoregressive Minimum Bayes' Risk Decoding \\
for Fast Speech Recognition
}

\author[affiliation={1}, orcid=0000-0003-2127-6607]{Hiroyuki}{Deguchi}
\author[affiliation={1}, orcid=0009-0007-1291-8416]{Takatomo}{Kano}
\author[affiliation={1}, orcid=0009-0001-3197-3388]{Katsuki}{Chousa}
\author[affiliation={1}, orcid=0000-0002-5175-7834]{Marc}{Delcroix}

\address{
    $^1$ NTT, Inc., Japan
}

\email{
hiroyuki.deguchi@ntt.com, takatomo.kanou@ntt.com, \\
katsuki.chousa@ntt.com, marc.delcroix@ntt.com
}


\usepackage{comment}

\begin{document}

\maketitle

\begin{abstract}
Non-autoregressive (NAR) decoding generates output tokens in parallel, making speech recognition faster than autoregressive decoding, which generates them sequentially from left to right.
However, the recognition performance is degraded because NAR decoding cannot resolve uncertainty by conditioning on previously generated tokens.
To address this issue, we propose a novel NAR decoding framework based on minimum Bayes' risk (MBR) decoding, termed NAR-MBR decoding, that maximizes the expected utility calculated from samples drawn from the output probability of an NAR model rather than maximizing the output probability.
Notably, by leveraging the nature of NAR models, multiple samples are obtained efficiently with a single forward computation.
Our experiments across LibriSpeech, Switchboard, AMI, and web presentation corpus demonstrated that our NAR-MBR decoding outperformed previous NAR decoding and ran faster than AR decoding.
\end{abstract}

\section{Introduction}
\label{sec:intro}
Decoding speed in automatic speech recognition (ASR) is challenging in real-world scenarios, especially when processing long speech signals.
Most of the recent state-of-the-art ASR models that have achieved low word error rates (WER) generate transcriptions in a left-to-right manner, called autoregressive (AR) decoding, which takes computational time proportional to the number of output tokens~\cite{gulati-etal-2020-conformer,kim-etal-2022-e-branchformer,radford-etal-2023-robust,puvvada-etal-2024-less}.
Non-autoregressive (NAR) decoding addresses the limitations of AR decoding by introducing an independence assumption and removing the dependency on previously generated tokens in the output probability, i.e., context information~\cite{gu-etal-2018-nonautoregressive,libovicky-helcl-2018-end,ghazvininejad-etal-2019-mask,higuchi-etal-2020-mask}.
This enables multiple tokens to be generated in parallel, and the number of decoding steps is reduced to either one or a constant smaller than the length of the output sequence, making it faster than AR decoding.
Nevertheless, there remains a performance gap between AR and NAR decoding due to the multi-modality problem in NAR decoding~\cite{gu-etal-2018-nonautoregressive}, which makes it difficult to maintain consistency in probable path selection when transcriptions may have uncertainty among multiple output paths.

In the field of decision theory, the expected utility theory (EUT)~\cite{von-neumann-morgenstern-1944-theory} is often used for decision-making under uncertainty.
In EUT, optimal decisions maximize the expected utility (EU), where the utility function represents preference relations or desirability.
The minimum Bayes' risk (MBR) framework based on EUT\footnote{
EU maximization is equivalent to Bayes' risk minimization.
} has been widely utilized for both training~\cite{doumpiotis-etal-2003-discriminative,doumpiotis-etal-2003-lattice,prabhavalkar-etal-2018-minimum,kanda-etal-2021-minimum,tian-etal-2023-bayes} and decoding~\cite{goel-and-byrne-2000-minimum,kumar-byrne-2004-minimum,shafran-and-byrne-2004-task,xu-etal-2010-improved,eikema-aziz-2020-map} in ASR and other fields.
In particular, MBR decoding improves the recognition performance without additional training by selecting the output sequence that maximizes the EU, in contrast to the most widely used decoding method, maximum a posteriori (MAP) decoding, which selects the sequence that maximizes the output probability.
The EU is typically estimated using the Monte Carlo method, with multiple output samples drawn from the output probability distribution.

\begin{figure*}
    \begin{tabular}{@{}cc@{}}
    \begin{minipage}{0.48\linewidth}
        \centering
        \includegraphics[width=\linewidth]{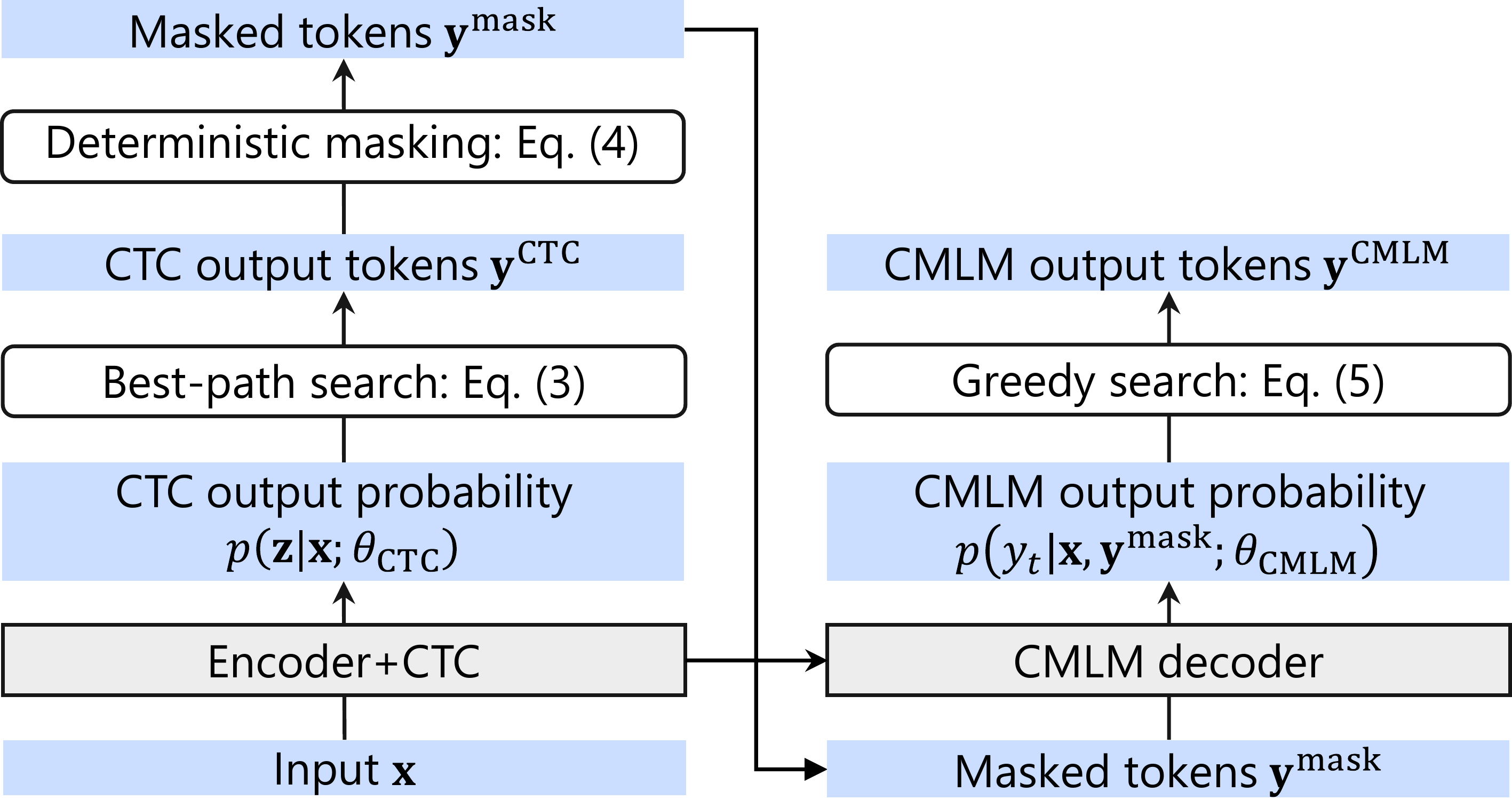}
        \subcaption{NAR decoding in Mask-CTC}
        \label{fig:overview:nar}
    \end{minipage}
    &
    \begin{minipage}{0.48\linewidth}
        \centering
        \includegraphics[width=\linewidth]{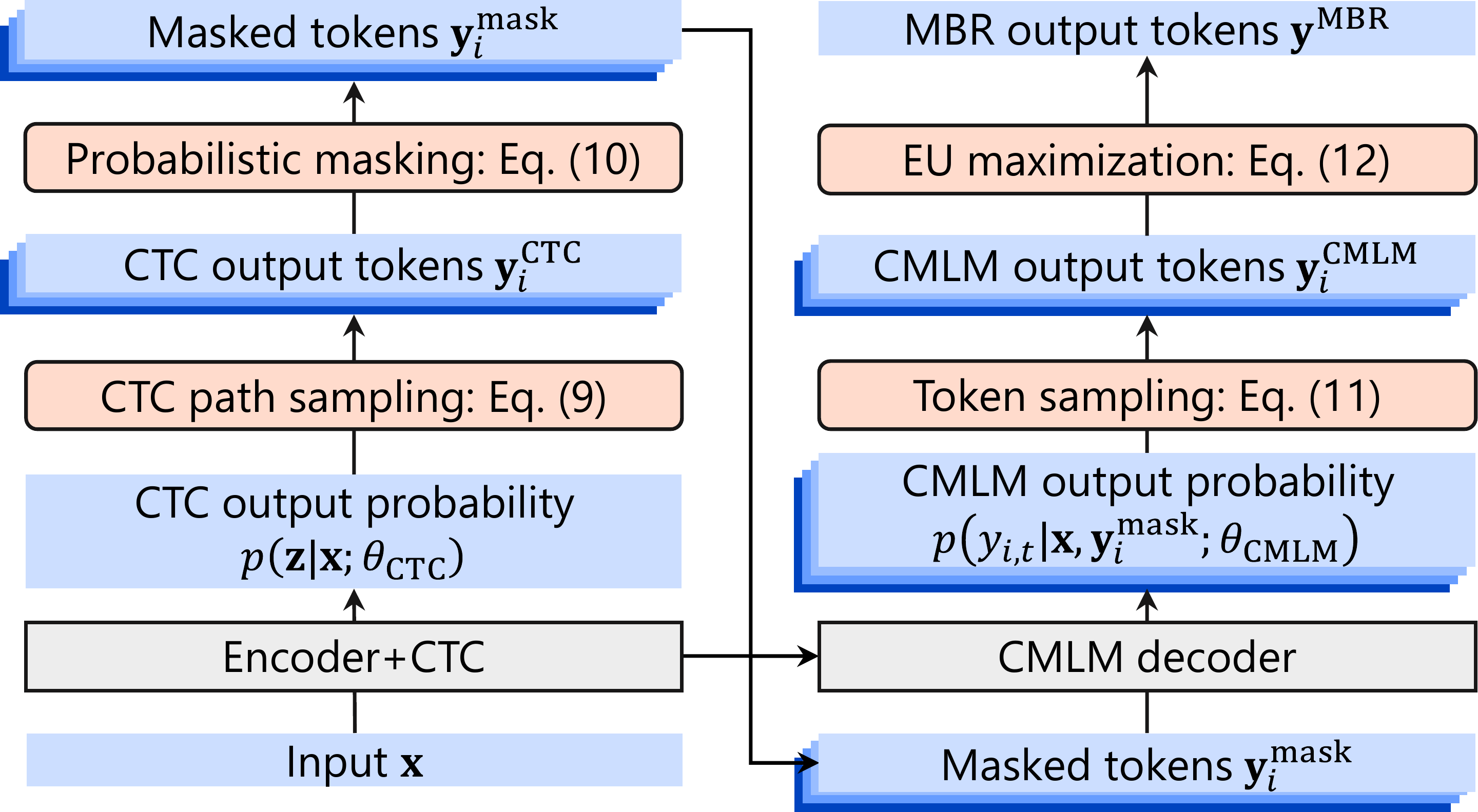}
        \subcaption{Proposed NAR-MBR decoding in Mask-CTC}
        \label{fig:overview:narmbr}
    \end{minipage}
    \end{tabular}
    \caption{NAR and NAR-MBR decoding in Mask-CTC at $\niter=1$.
    Differences from NAR decoding are \colorbox[HTML]{FCDBCD}{highlighted}.
    }
    \label{fig:overview}
\end{figure*}

For fast yet high-quality speech recognition, we propose NAR-MBR decoding that maximizes the EU by using efficient non-autoregressive sampling.
\zcref{fig:overview} presents an overview of NAR and our NAR-MBR decoding.
NAR-MBR decoding generates multiple output samples using a non-autoregressive ASR model and then maximizes the EU based on WER.
For estimating the EU, we generate multiple samples drawn from the output probability distributions of an NAR model utilizing unbiased sampling rather than greedy search.
Notably, we efficiently sample multiple output paths only with a single forward computation by leveraging the independence assumption between output tokens in NAR models.
We then estimate and maximize the EU using our efficient edit distance computation.

Experiments on LibriSpeech~\cite{panayotov-etal-2015-librispeech}, Switchboard~\cite{godfrey-etal-1992-switchboard}, AMI~\cite{mccowan-etal-2005-ami}, and web presentation corpus showed that NAR-MBR decoding transcribed speech up to 43.1 times faster than the beam search of AR decoding using Conformer~\cite{gulati-etal-2020-conformer} and achieved both speed and accuracy improvements, without additional training, compared to NAR decoding using Mask-CTC~\cite{higuchi-etal-2020-mask}.

\section{Background: Decoding in ASR}
\label{sec:bg}

Let $\spaceIn$ be the input space of ASR models and $\spaceOut \coloneqq \vcbTok^\ast$ be the output space, where $\vcbTok^\ast$ is the Kleene closure of vocabulary $\vcbTok$.
The goal of ASR is to generate a token sequence $\seqOut \in \spaceOut$ from an input speech $\seqIn \in \spaceIn$.
Output tokens $\seqOut$ are converted into the transcription text $\funcDetok(\seqOut) \in \vcbAlph^\ast$, where $\vcbAlph^\ast$ is the Kleene closure of alphabet $\vcbAlph$, and $\funcDetok\colon \vcbTok^\ast \to \vcbAlph^\ast$ denotes a detokenizing function that maps a token sequence to its corresponding string.

\subsection{Sequence modeling in decoding}
\label{sec:bg:nar}

\parag{Autoregressive decoding}
Most ASR models are trained to calculate the output probability of $\seqOut$ given $\seqIn$, i.e., $p(\seqOut | \seqIn; \param)$, where $\param$ denotes trained parameters.
Typically, they find the most probable path that maximizes the output probability, 
which is formulated using the chain rule of conditional probabilities:
\begin{equation}
    p_\decAr(\seqOut | \seqIn; \theta) \coloneqq \textstyle\prod_{\curTime=1}^{|\seqOut|} p(\tokOut_\curTime | \seqIn, \seqOut_{<\curTime}).
\end{equation}
In this autoregressive manner, output probabilities are predicted sequentially from left to right, conditioned on previously generated tokens, allowing ambiguities among paths to be resolved.
However, it is time-consuming, especially for long speech signals, because probabilities must be calculated $|\seqOut|$ times.

\parag{Non-autoregressive decoding}
For faster decoding, NAR decoding reduces the number of probability calculations.
It generates multiple tokens in parallel, based on the independence assumption between output tokens, as follows:
\begin{equation}
    p_\decNar(\seqOut | \seqIn; \theta) \coloneqq \textstyle\prod_{\curTime=1}^{|\seqOut|} p(\tokOut_\curTime | \seqIn).
\end{equation}
Mask-CTC~\cite{higuchi-etal-2020-mask}, one of the most popular non-autoregressive ASR models, is built on an encoder--decoder architecture (\zcref{fig:overview:nar}), where the encoder concurrently generates the entire output sequence with connectionist temporal classification (CTC)~\cite{graves-etal-2006-connectionist,libovicky-helcl-2018-end}, and the decoder, based on a conditional masked language model (CMLM)~\cite{ghazvininejad-etal-2019-mask}, refines the encoder outputs with mask prediction.
The encoder predicts the output probability of tokens and blanks and greedily selects the CTC alignment path that maximizes the probability for each frame, as
\begin{equation}
    \seqCtc^\modelCtc \coloneqq \argmax_{\seqCtc \in \vcbTok_\modelCtc^\ast} p(\seqCtc | \seqIn; \paramCtc{}) = \argmax_{\seqCtc \in \vcbTok_\modelCtc^\ast} \textstyle\prod_{\curTime=1}^{|\seqCtc|} p(\tokCtc_\curTime | \seqIn; \paramCtc),
\end{equation}
where $\vcbTok_\modelCtc \coloneqq \vcbTok \cup \{\blank\}$ is the CTC output vocabulary of the encoder and $\blank$ denotes a blank frame.
CTC then collapses the alignment path $\seqCtc^\modelCtc$ by removing all blanks and aggregating consecutive frames with the same tokens.
Let $\funcCtcPath\colon \vcbTok_\modelCtc^\ast \twoheadrightarrow \spaceOut$ be a surjective function that maps a CTC path to its corresponding collapsed output sequence.
The CTC output sequence is defined as $\seqOut^\modelCtc \coloneqq \funcCtcPath(\seqCtc^\modelCtc)$.
The CMLM decoder masks tokens within the CTC output $\seqOut^\modelCtc$ according to its confidence probability $p(\tokOut^\modelCtc_\curTime | \seqIn; \paramCtc)$, and fills the masked tokens.
The masks $\seqMask \in \{ 0, 1 \}^\ast$ are determined with a threshold $\thresMask \in [0, 1]$:
\begin{equation}
    \tokMask_\curTime \coloneqq \mathbbm{1}[p(\tokOut^\modelCtc_\curTime | \seqIn; \paramCtc) < \thresMask],
\end{equation}
where $\mathbbm{1}$ denotes an indicator function.
The masked sequence $\seqOut^\text{mask}$ is obtained by masking the CTC output using $\seqMask$:
if $\tokMask_\curTime=1$, $\tokOut_\curTime$ is replaced with the mask token; otherwise, it remains unchanged.
Mask tokens are then filled via mask prediction:
\begin{equation}
    \tokOut^\modelCmlm_\curTime \!\coloneqq
    \begin{cases}
        \textstyle\argmax_{\tokOut_\curTime \in \vcbTok} p(\tokOut_\curTime | \seqIn, \seqOut^\text{mask}; \paramCmlm) &  \tokMask_\curTime = 1 \\
        \tokOut^\text{mask}_\curTime & \tokMask_\curTime = 0
    \end{cases}.
\end{equation}
The mask prediction can also be split into $\niter \in \Z_{\geq 0}$ inference stages instead of filling all mask tokens at once, at the expense of decoding speed.
Note that $\niter = 0$ corresponds to using $\seqOut^\modelCtc$ without the CMLM.
By filling tokens starting with those of higher confidence, lower WER has been achieved~\cite{higuchi-etal-2020-mask}.

Mask-CTC corrects erroneous tokens that occur due to the non-autoregressive manner of the encoder with CTC, and achieves a better recognition performance than encoder-only NAR models while reducing the number of probability calculations and generating outputs faster than AR decoding.
Nevertheless, there is still a performance gap between AR and NAR decoding due to the multi-modality problem~\cite{gu-etal-2018-nonautoregressive} caused by the uncertainty of multiple output paths.

\subsection{Decision rules in decoding}
\label{sec:bg:mbr}
\parag{MAP decoding}
The most widely used output decision rule in ASR is MAP decoding, which maximizes the output probability.
Since the output space $\spaceOut$ is an infinite set and searching over all possible paths is infeasible, it selects a pruned hypothesis set $\setHyp \subset \spaceOut$, usually obtained via beam search or sampling:
\begin{equation}
    \label{eq:map}
    \seqOut^\decMap \coloneqq \textstyle\argmax_{\seqHyp \in \setHyp} p(\seqHyp | \seqIn; \param).
\end{equation}
This point estimation strategy can pose problems, and recent studies have shown that high-probability sequences are not always high-quality.
Specifically, as the beam size increases, sequences with higher probability are generated.
However, such sequences hurt output quality and sometimes cause pathological outputs, e.g., empty outputs and $n$-gram repetitions~\cite{koehn-knowles-2017-six,ott-etal-2018-analyzing}.

\parag{MBR decoding}
As a more quality-aware decision rule, MBR decoding selects the hypothesis $\seqOut^\decMbr$ that maximizes the EU:
\begin{equation}
    \label{eq:mbr}
    \seqOut^\decMbr \coloneqq \textstyle\argmax_{\seqHyp \in \setHyp} \expect_{\seqOut \sim \Pr(\cdot | \seqIn)} [\funcUtility(\seqHyp; \seqOut)],
\end{equation}
where $\Pr(\cdot|\seqIn)$ is the true output probability distribution given the input $\seqIn$,
and $\funcUtility\colon \spaceOut \times \spaceOut \to \R$ is the utility function,
which satisfies $\seqHyp \succeq_\seqOut \seqHyp' \,\iff\, \funcUtility(\seqHyp; \seqOut) \geq \funcUtility(\seqHyp'; \seqOut)$,
where $\succeq_\seqOut$ denotes the preference relation under the given reference $\seqOut$.
Typically, evaluation metrics to be maximized, e.g., negative WER, are used for the utility function.
Since $\Pr(\cdot|\seqIn)$ is unknown, the EU is commonly estimated with the Monte Carlo (MC) method using output samples drawn from the model $\param$~\cite{eikema-aziz-2020-map,eikema-aziz-2022-sampling}:
\begin{equation}
    \label{eq:mcmbr}
    \seqOut^\decMbr \simeq \textstyle\argmax_{\seqHyp \in \setHyp} 
    \frac{1}{|\setPref|} \textstyle\sum_{\seqOut \in \setPref} \funcUtility(\seqHyp; \seqOut),
\end{equation}
where $\setPref \coloneqq \{\seqOut_i \}_{i=1}^{|\setPref|} \overset{\mathrm{i.i.d.}}{\sim} p(\cdot|\seqIn; \param)$ is a multiset of samples, called pseudo-references, which are drawn from the output probability distribution.
To estimate the EU stably, unbiased sampling methods are used for generating pseudo-references rather than beam search or other biased sampling methods~\cite{eikema-aziz-2022-sampling,freitag-etal-2023-epsilon}.
In a typical setting, the same sample set is used for both the hypotheses and the pseudo-references.
MBR decoding robustly generates high-quality outputs but is computationally expensive due to hypothesis and pseudo-reference sampling and the quadratic-time utility calculation.

\section{Proposed Method: NAR-MBR Decoding}
Our non-autoregressive minimum Bayes' risk (NAR-MBR) decoding aims to decode high-quality transcriptions significantly faster than AR decoding, without additional training for Mask-CTC.
\zcref{fig:overview:narmbr} presents the overview of NAR-MBR decoding.
It mainly consists of two steps:
(1) probabilistic unbiased sampling from Mask-CTC, and (2) efficient EU maximization.

\subsection{Probabilistic unbiased sampling from Mask-CTC}
\label{sec:prop:sampling}
We sample hypotheses and pseudo-references from the output probability distribution of Mask-CTC.
As described in \zcref{sec:bg:mbr}, MBR decoding requires unbiased sampling to generate pseudo-references, which is usually computationally expensive.
While Mask-CTC generates CTC paths via fast but biased greedy search, we instead perform unbiased sampling without additional cost to obtain multiple CTC alignment paths.

We first sample multiple CTC paths.
Notably, this step does not increase the computational cost.
In AR models, sequence sampling increases the computational cost by the number of samples, as it calculates context-dependent conditional probabilities.
Here, we sample sequences from NAR models, i.e., from context-independent probabilities.
Since the probabilities are computed in a single forward pass, multiple samples can be obtained efficiently.
The paths are sampled independently according to the output probabilities for each frame:
\begin{equation}
\label{eq:sample:ctc}
    \setSampleCtc \coloneqq \{ \seqCtc_i \}_{i=1}^{|\setSampleCtc|} \overset{\mathrm{i.i.d.}}{\sim}
    p(\seqCtc | \seqIn; \paramCtc), ~~~
    \tokCtc_{i, t} \sim \distCategorical{ p(\tokCtc_\curTime | \seqIn; \paramCtc)},
\end{equation}
where $\setSampleCtc$ is a multiset of sampled CTC alignment paths and $\mathrm{Cat}(\cdot)$ denotes a categorical distribution.
We then perform mask prediction for each $\seqCtc_i \in \setSampleCtc$.
The masks are sampled from Bernoulli distributions based on the confidence probabilities:
\begin{align}
    m_{i,\curTime} &\sim \distBernoulli{ 1 - p(\tokOut^\modelCtc_{i,\curTime} | \seqIn; \paramCtc) }.
\end{align}
Unconfident tokens are masked according to $\seqMask_i$, the same as in the previous method.
We then fill in the masked tokens based on the categorical distribution of the decoder's output probabilities.
\begin{align}
    \tokOut^\modelCmlm_{i,\curTime} \coloneqq 
    \begin{cases}
        \tokOut_{i,\curTime} \sim \distCategorical{p( \tokOut_t | \seqIn, \seqOut^\text{mask}_i; \paramCmlm)} & \tokMask_{i,\curTime} = 1 \\
        \tokOut^\text{mask}_{\curTime} & \tokMask_{i,\curTime} = 0
    \end{cases}.
\end{align}
Similar to standard decoding in Mask-CTC, our method can also extend mask prediction to $\niter$ inference stages.
To probabilistically fix output tokens from higher confident ones, we use the Gumbel-max trick~\cite{gumbel0-1954-statistical} instead of top-k.

Using the above method, a sample set for pseudo-references $\setPref$ is obtained.
Following previous studies~\cite{eikema-aziz-2020-map,eikema-aziz-2022-sampling}, we use the same set for both hypotheses $\setHyp$ and pseudo-references $\setPref$.
For $\niter=0$, CTC path samples $\setSampleCtc$ are directly used as $\setHyp$ and $\setPref$.

\subsection{Efficient EU maximization}
\label{sec:prop:mbr}
We select the best sequence that maximizes the EU from the hypotheses based on \zcref{eq:mcmbr}.
For the utility function $\funcUtility$, we employ negative WER, i.e., $\funcUtility(\seqHyp; \seqOut) = -\funcWer(\funcDetok(\seqHyp); \funcDetok(\seqOut))$, where $\funcWer\colon \vcbAlph^\ast \times \vcbAlph^\ast \to [0, \infty)$ calculates a WER score given a reference transcription $\funcDetok(\seqOut)$.
The output is thus obtained as
\begin{align}
    \label{eq:mcmbr:asr}
    \seqOut^\decMbr \simeq \textstyle\argmax_{\seqHyp \in \setHyp} 
    -\frac{1}{|\setPref|} \textstyle\sum_{\seqOut \in \setPref} \funcWer(\funcDetok(\seqHyp); \funcDetok(\seqOut)).
\end{align}
Estimating the EU for all hypotheses requires a quadratic time proportional to the number of samples.
To maintain the speed benefit gained by using NAR models, 
we efficiently calculate the EU and mitigate the speed reduction due to MBR decoding.

\parag{Removing longest common prefix and suffix}
Removing the longest common prefix and suffix between a hypothesis and a pseudo-reference does not affect the edit distance.
We thus remove them before the calculation to reduce sequence lengths.

\parag{Memorization}
By caching the scores of duplicate sample pairs, we reduce the number of utility function calls.
Since pseudo-references are sampled as a multiset, it sometimes contains duplicate samples.
We first extract unique hypotheses and pseudo-references, calculate and cache the occurrence counts and scores of unique hypothesis--pseudo-reference pairs, and reuse the cached results for duplicate pairs.

\parag{Parallelization}
We calculate scores for each unique pair in parallel by leveraging multiple CPUs.

\parag{Implementation}
We implement WER in Rust, including text normalization.
In the edit distance calculation, we convert words to \texttt{u32} word IDs to avoid expensive string comparisons.

\section{Experiments}
\subsection{Setup}
To confirm the effectiveness of NAR-MBR decoding, we evaluate recognition performance and decoding speed in ASR tasks.

\parag{Datasets}
We use LibriSpeech~\cite{panayotov-etal-2015-librispeech}, 
Switchboard (SWBD)~\cite{godfrey-etal-1992-switchboard},
AMI~\cite{mccowan-etal-2005-ami}, 
and web presentation corpus (Web).
The Web corpus consists of 346 hours of training data from 1,938 speakers, and 3.7 hours of development and test sets, each from 16 speakers.
LibriSpeech consists of ``Clean'' and ``Other'', and SWBD consists of ``Switchboard (Swbd)'' and ``Callhome (Callhm).''

\parag{Models}
We use Conformer~\cite{gulati-etal-2020-conformer} for AR decoding and Mask-CTC~\cite{higuchi-etal-2020-mask} for both NAR and NAR-MBR decoding.
We train the models with the default hyperparameters for each dataset, as defined in \toolEspnet{}~\cite{watanabe-etal-2018-espnet}.
The only differences between the models are the architectures and loss functions, i.e., the number of model parameters and other hyperparameters are the same.

\parag{Decoding}
For AR decoding, we use joint decoding with CTC~\cite{kim-etal-2017-joint-ctc,hori-etal-2017-joint} and set the CTC weight to 0.3.
We compare the beam widths of 1 (Greedy) and 10 (Beam).
For NAR decoding, we use $\thresMask = 0.999$~\cite{higuchi-etal-2020-mask} as the masking threshold.
In NAR-MBR decoding, we compare the sampling sizes $|\setSampleCtc| \in \{ 64, 256 \}$ in \zcref{eq:sample:ctc}.
In NAR and NAR-MBR decoding, we compare the number of inference iterations $\niter \in \{ 0, 1, 10 \}$, where $\niter = 0$ directly generates transcriptions from the CTC outputs without the CMLM decoder.

\parag{Evaluation metrics}
We evaluate recognition performance using WER.
We also conduct a statistical significance test using paired bootstrap resampling~\cite{koehn-2004-statistical,bisani-and-ney-2004-bootstrap} with 1,000 resamples, comparing NAR-MBR decoding with the baseline NAR decoding.

\parag{Efficiency evaluation}
For evaluating decoding speed, we calculate the gain for each decoding method relative to Beam decoding (Speedup).
Specifically, we measure the total wall-clock time for forward computation in the encoder and decoder, including utility calculation time in NAR-MBR decoding.
We do not include other processing times, such as model and data loading, in the measurement.
In addition, we measure the ratio of average GPU memory usage relative to Beam decoding (Mem).

\parag{Computational environments}
We use 8 cores of Intel\regmark{} Xeon\regmark{} Gold 6346 CPU @ 3.10GHz and an NVIDIA\regmark{} RTX\trademark{} 6000 Ada GPU.
In all experiments, we employ \toolEspnet~\cite{watanabe-etal-2018-espnet} for training models and \toolMbrs~\cite{deguchi-etal-2024-mbrs} for MBR decoding.

\subsection{Results}
\begin{table}[t]
    \centering
    \footnotesize
    \caption{Comparison of WER for each decoding method.
    \textbf{Bold} and \underline{underlined} texts indicate the best and second-best WER, respectively, among NAR and NAR-MBR decoding.
    ``$\dagger$'' indicates that NAR-MBR decoding significantly outperforms NAR decoding for all $\niter \in \{ 0, 1, 10\}$ ($p<0.05$).
    }
    \label{tab:results:wer}
    \begin{NiceTabular}{@{}l rr rr rr@{}}
        \toprule
                & \multicolumn{2}{@{}c@{}}{LibriSpeech} & \multicolumn{2}{@{}c@{}}{SWBD} & & \\
        \cmidrule(lr){2-3}
        \cmidrule(l){4-5}
         Decoding 
         & Clean & Other
         & Swbd & Callhm
         & AMI
         & Web
         \\
         \midrule
         \multicolumn{6}{@{}l@{}}{
            \rowcolor{gray!20}
            AR
         }
         \\
         Greedy & 3.0 & 6.0 & 6.9 & 13.9 & 17.8 & 8.2 \\
         Beam   & 2.4 & 5.5 & 6.6 & 13.5 & 17.0 & 7.3 \\
         \midrule
         \multicolumn{6}{@{}l@{}}{
            \rowcolor{gray!20}
            NAR
         }
         \\
         $\niter=0$  & 3.3 & 7.4 & 7.9 & 15.7 & 18.9 & 7.7 \\
         $\niter=1$  & 3.4 & 7.7 & 7.8 & 15.6 & 18.8 & 8.9 \\
         $\niter=10$ & 3.3 & 7.5 & 7.6 & \underline{15.2} & \underline{18.4} & 8.5 \\
         \multicolumn{6}{@{}l@{}}{
            \rowcolor{gray!20}
            NAR-MBR with $|\setSampleCtc| = 64$ (ours)
         }
         \\
         $\niter=0$ & 3.3 & 7.4 & 7.8 & 15.6 & 18.7 & ${}^\dagger$7.6 \\
         $\niter=1$ & ${}^\dagger$\textbf{3.1} & ${}^\dagger$\underline{7.1} & ${}^\dagger$\textbf{7.3} & ${}^\dagger$\textbf{14.9} & ${}^\dagger$\textbf{18.1} & ${}^\dagger$\underline{7.4} \\
         $\niter=10$ & ${}^\dagger$\textbf{3.1} & ${}^\dagger$\underline{7.1} & ${}^\dagger$\underline{7.4} & ${}^\dagger$\textbf{14.9} & ${}^\dagger$\textbf{18.1} & ${}^\dagger$\underline{7.4} \\
         \multicolumn{6}{@{}l@{}}{
            \rowcolor{gray!20}
            NAR-MBR with $|\setSampleCtc| = 256$ (ours)
         }
         \\
         $\niter=0$  & \underline{3.2} & 7.4 & 7.7 & 15.5 & 18.6 & ${}^\dagger$7.5 \\
         $\niter=1$  & ${}^\dagger$\textbf{3.1} & ${}^\dagger$\textbf{7.0} & ${}^\dagger$\textbf{7.3} & ${}^\dagger$\textbf{14.9} & ${}^\dagger$\textbf{18.1} & ${}^\dagger$\textbf{7.3} \\
         $\niter=10$ & ${}^\dagger$\textbf{3.1} & ${}^\dagger$\textbf{7.0} & ${}^\dagger$\textbf{7.3} & ${}^\dagger$\textbf{14.9} & ${}^\dagger$\textbf{18.1} & ${}^\dagger$\textbf{7.3} \\
         \bottomrule
    
    \end{NiceTabular}
\end{table}

\begin{table}[t]
    \centering
    \footnotesize
    \tabcolsep 4.5pt
    \caption{Decoding speed and average GPU memory usage on LibriSpeech (LS) and Web, relative to Beam.
    Results of $\niter=10$ in NAR-MBR decoding are omitted since peak WER was achieved at $\niter=1$,
    as shown in \zcref{tab:results:wer}.
    }
    \label{tab:results:speed}
    \begin{NiceTabular}{@{}l rr rr rr@{}}
    \toprule
         & \multicolumn{2}{@{}c@{}}{LS (Clean)}
         & \multicolumn{2}{@{}c@{}}{LS (Other)}
         & \multicolumn{2}{@{}c@{}}{Web}
         \\
         \cmidrule(lr){2-3}
         \cmidrule(lr){4-5}
         \cmidrule(lr){6-7}

        Decoding
         & Speed$\uparrow$ & Mem$\downarrow$
         & Speed$\uparrow$ & Mem$\downarrow$
         & Speed$\uparrow$ & Mem$\downarrow$
         \\
         \midrule
         \multicolumn{6}{@{}l@{}}{
            \rowcolor{gray!20}
            AR
         }
         \\
         Greedy & $\times$5.3 & $\times$1.0 & $\times$5.2 & $\times$1.0 & $\times$5.0 & $\times$1.0 \\
         Beam   & $\times$1.0 & $\times$1.0 & $\times$1.0 & $\times$1.0 & $\times$1.0 & $\times$1.0 \\
         \midrule
         \multicolumn{6}{@{}l@{}}{
            \rowcolor{gray!20}
            NAR
         }
         \\
         $\niter=0$  & $\times$61.3 & $\times$1.0 & $\times$50.1 & $\times$1.0 & $\times$90.3 & $\times$1.0 \\
         $\niter=1$  & $\times$44.2 & $\times$1.0 & $\times$34.7 & $\times$1.0 & $\times$71.3 & $\times$1.0 \\
         $\niter=10$ & $\times$21.3 & $\times$1.0 & $\times$15.2 & $\times$1.0 & $\times$26.7 & $\times$1.0\\
         \multicolumn{6}{@{}l@{}}{
            \rowcolor{gray!20}
            NAR-MBR with $|\setSampleCtc| = 64$ (ours)
         }
         \\
         $\niter=0$  & $\times$38.7 & $\times$1.0 & $\times$32.1 & $\times$1.0 & $\times$75.2 & $\times$1.0 \\
         $\niter=1$  & $\times$27.4 & $\times$1.3 & $\times$22.4 & $\times$1.3 & $\times$43.1 & $\times$1.8 \\
         \multicolumn{6}{@{}l@{}}{
            \rowcolor{gray!20}
            NAR-MBR with $|\setSampleCtc| = 256$ (ours)
         }
         \\
         $\niter=0$  & $\times$30.9 & $\times$1.0 & $\times$22.1 & $\times$1.0 & $\times$41.1 & $\times$1.0\\
         $\niter=1$  & $\times$11.8 & $\times$2.7 & $\times$9.7 & $\times$2.4 & $\times$20.7 & $\times$5.0 \\
         \bottomrule
    \end{NiceTabular}
    
\end{table}

\parag{Recognition performance}
\zcref{tab:results:wer} lists the WER for each decoding method.
In all datasets, NAR-MBR decoding consistently outperformed NAR decoding for both $|\setSampleCtc| = 64$ and $256$ when $\niter \in \{ 1, 10\}$, with statistically significant improvements.
Notably, $\niter=1$ significantly improved WER compared to NAR decoding and achieved a comparable performance to beam search in the Web corpus.

In the comparisons within NAR-MBR decoding, we observed two trends.
First, increasing the sample size improved the WER, slightly.
This is because the EU is estimated more robustly as the number of pseudo-references increases, owing to the nature of the Monte Carlo estimation in MBR decoding.
Second, a larger number of inference iterations $\niter$ did not improve recognition performance.
Interestingly, the WER of NAR-MBR decoding converged
at $\niter=1$ and did not improve with a larger $\niter$, while NAR decoding did so.
In NAR decoding, as the number of CMLM decoder calls increases, the condition of mask prediction becomes richer, making it easier to resolve ambiguities.
In contrast, in NAR-MBR decoding, EU maximization serves this role instead.
In addition, EU estimation requires unbiased samples rather than the most probable output, and we can obtain a sufficient number of such samples with a single forward computation.
Therefore, NAR-MBR decoding achieved low WER without iterative refinement.

To summarize, we observed that NAR-MBR decoding outperformed NAR decoding and peak performance was achieved at $\niter=1$ for both $|\setSampleCtc| = 64$ and $256$.

\parag{Decoding speed}
\zcref{tab:results:speed} shows the comparisons of decoding speed and average GPU memory usage.
We omitted the results of $\niter=10$ in NAR-MBR decoding since peak WER was achieved at $\niter=1$, as shown in \zcref{tab:results:wer}.
The results demonstrate that NAR-MBR decoding consistently ran faster than AR decoding, including Greedy decoding.
In particular, in the Web corpus, NAR-MBR decoding at $\niter=1$ ran 43.1 and 20.7 times faster than Beam decoding in $|\setSampleCtc| = 64$ and $256$, respectively, while maintaining a comparable WER.
Moreover, compared to NAR decoding at $\niter=10$, NAR-MBR decoding with 64 samples at $\niter=1$ was faster, yet it significantly improved the WER.
The memory usage of NAR-MBR decoding was comparable to that of other methods when $\niter=0$ but increased when $\niter=1$, primarily due to the computational cost of the CMLM decoder.
Addressing this issue through more memory-efficient sampling will be part of future work.

\subsection{Effect of the number of samples}

\begin{figure}[t]
    \centering
    \includegraphics[width=1.0\linewidth]{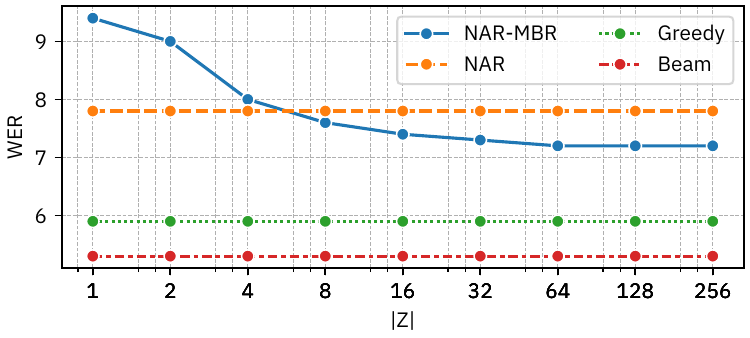}
    \caption{
    WER of NAR-MBR decoding at $\niter=1$ when varying number of samples
    in dev. set (Other) of LibriSpeech
    }
    \label{fig:nsamples:libri}
\end{figure}

We investigate the effect of the number of samples $|\setSampleCtc|$ on the recognition performance in NAR-MBR decoding.
\zcref{fig:nsamples:libri} shows the WER when varying the number of samples within $|\setSampleCtc| \in \{ 2^0, \ldots, 2^8 \}$ in the development set of LibriSpeech (Other).
The results show that NAR-MBR decoding improved the WER as the number of samples increased.
In addition, the gain of NAR-MBR decoding converged in $|\setSampleCtc| \geq 64$ and showed a similar trend to existing theoretical analyses of convergence rate in autoregressive MBR decoding~\cite{ichihara-etal-2025-theoretical,kamigaito-etal-2025-diversity}.

\section{Conclusion}
This paper proposes NAR-MBR decoding for fast speech recognition.
The proposed method addresses the performance degradation of NAR decoding via probabilistic sampling and EU maximization.
Experimental results demonstrated that our NAR-MBR decoding outperforms previous NAR decoding without additional training, and achieves a better trade-off between recognition performance and decoding speed.
For future work, we will extend our approach to other models and tasks.

\section{Generative AI use disclosure}
Our use of generative AI tools was limited to proofreading, such as grammar and spell-checkers.
All implementations, experiments, and analyses were conducted by the authors, who take full responsibility for the content.

\bibliographystyle{IEEEtran}
\bibliography{mybib,anthology-1,anthology-2}

\begin{thebibliography}{10}
\providecommand{\url}[1]{#1}
\csname url@samestyle\endcsname
\providecommand{\newblock}{\relax}
\providecommand{\bibinfo}[2]{#2}
\providecommand{\BIBentrySTDinterwordspacing}{\spaceskip=0pt\relax}
\providecommand{\BIBentryALTinterwordstretchfactor}{4}
\providecommand{\BIBentryALTinterwordspacing}{\spaceskip=\fontdimen2\font plus
\BIBentryALTinterwordstretchfactor\fontdimen3\font minus \fontdimen4\font\relax}
\providecommand{\BIBforeignlanguage}[2]{{%
\expandafter\ifx\csname l@#1\endcsname\relax
\typeout{** WARNING: IEEEtran.bst: No hyphenation pattern has been}%
\typeout{** loaded for the language `#1'. Using the pattern for}%
\typeout{** the default language instead.}%
\else
\language=\csname l@#1\endcsname
\fi
#2}}
\providecommand{\BIBdecl}{\relax}
\BIBdecl

\bibitem{gulati-etal-2020-conformer}
A.~Gulati, J.~Qin, C.-C. Chiu, N.~Parmar, Y.~Zhang, J.~Yu, W.~Han, S.~Wang, Z.~Zhang, Y.~Wu, and R.~Pang, ``{Conformer: Convolution-augmented Transformer for Speech Recognition},'' in \emph{{Interspeech 2020}}, 2020, pp. 5036--5040.

\bibitem{kim-etal-2022-e-branchformer}
K.~Kim, F.~Wu, Y.~Peng, J.~Pan, P.~Sridhar, K.~J. Han, and S.~Watanabe, ``E-branchformer: Branchformer with enhanced merging for speech recognition,'' in \emph{2022 IEEE Spoken Language Technology Workshop (SLT)}, 2023, pp. 84--91.

\bibitem{radford-etal-2023-robust}
\BIBentryALTinterwordspacing
A.~Radford, J.~W. Kim, T.~Xu, G.~Brockman, C.~Mcleavey, and I.~Sutskever, ``Robust speech recognition via large-scale weak supervision,'' in \emph{Proceedings of the 40th International Conference on Machine Learning}, ser. Proceedings of Machine Learning Research, A.~Krause, E.~Brunskill, K.~Cho, B.~Engelhardt, S.~Sabato, and J.~Scarlett, Eds., vol. 202.\hskip 1em plus 0.5em minus 0.4em\relax PMLR, 23--29 Jul 2023, pp. 28\,492--28\,518.
\BIBentrySTDinterwordspacing

\bibitem{puvvada-etal-2024-less}
K.~C. Puvvada, P.~Żelasko, H.~Huang, O.~Hrinchuk, N.~R. Koluguri, K.~Dhawan, S.~Majumdar, E.~Rastorgueva, Z.~Chen, V.~Lavrukhin, J.~Balam, and B.~Ginsburg, ``{Less is More: Accurate Speech Recognition \& Translation without Web-Scale Data},'' in \emph{{Interspeech 2024}}, 2024, pp. 3964--3968.

\bibitem{gu-etal-2018-nonautoregressive}
\BIBentryALTinterwordspacing
J.~Gu, J.~Bradbury, C.~Xiong, V.~O. Li, and R.~Socher, ``Non-autoregressive neural machine translation,'' in \emph{International Conference on Learning Representations}, 2018.
\BIBentrySTDinterwordspacing

\bibitem{libovicky-helcl-2018-end}
\BIBentryALTinterwordspacing
J.~Libovick{\'y} and J.~Helcl, ``End-to-end non-autoregressive neural machine translation with connectionist temporal classification,'' in \emph{Proceedings of the 2018 Conference on Empirical Methods in Natural Language Processing}, E.~Riloff, D.~Chiang, J.~Hockenmaier, and J.~Tsujii, Eds.\hskip 1em plus 0.5em minus 0.4em\relax Brussels, Belgium: Association for Computational Linguistics, Oct.-Nov. 2018, pp. 3016--3021.
\BIBentrySTDinterwordspacing

\bibitem{ghazvininejad-etal-2019-mask}
\BIBentryALTinterwordspacing
M.~Ghazvininejad, O.~Levy, Y.~Liu, and L.~Zettlemoyer, ``Mask-predict: Parallel decoding of conditional masked language models,'' in \emph{Proceedings of the 2019 Conference on Empirical Methods in Natural Language Processing and the 9th International Joint Conference on Natural Language Processing (EMNLP-IJCNLP)}, K.~Inui, J.~Jiang, V.~Ng, and X.~Wan, Eds.\hskip 1em plus 0.5em minus 0.4em\relax Hong Kong, China: Association for Computational Linguistics, Nov. 2019, pp. 6112--6121.
\BIBentrySTDinterwordspacing

\bibitem{higuchi-etal-2020-mask}
Y.~Higuchi, S.~Watanabe, N.~Chen, T.~Ogawa, and T.~Kobayashi, ``{Mask CTC: Non-Autoregressive End-to-End ASR with CTC and Mask Predict},'' in \emph{{Interspeech 2020}}, 2020, pp. 3655--3659.

\bibitem{von-neumann-morgenstern-1944-theory}
J.~von Neumann and O.~Morgenstern, \emph{Theory of Games and Economic Behavior}.\hskip 1em plus 0.5em minus 0.4em\relax Princeton: Princeton University Press, 1944.

\bibitem{doumpiotis-etal-2003-discriminative}
V.~Doumpiotis, S.~Tsakalidis, and W.~Byrne, ``Discriminative training for segmental minimum bayes risk decoding,'' in \emph{2003 IEEE International Conference on Acoustics, Speech, and Signal Processing, 2003. Proceedings. (ICASSP '03).}, vol.~1, 2003, pp. 136--139.

\bibitem{doumpiotis-etal-2003-lattice}
V.~Doumpiotis, S.~Tsakalidis, and W.~J. Byrne, ``{Lattice segmentation and minimum Bayes risk discriminative training},'' in \emph{{8th European Conference on Speech Communication and Technology (Eurospeech 2003)}}, 2003, pp. 1985--1988.

\bibitem{prabhavalkar-etal-2018-minimum}
R.~Prabhavalkar, T.~N. Sainath, Y.~Wu, P.~Nguyen, Z.~Chen, C.-C. Chiu, and A.~Kannan, ``Minimum word error rate training for attention-based sequence-to-sequence models,'' in \emph{2018 IEEE International Conference on Acoustics, Speech and Signal Processing (ICASSP)}, 2018, pp. 4839--4843.

\bibitem{kanda-etal-2021-minimum}
N.~Kanda, Z.~Meng, L.~Lu, Y.~Gaur, X.~Wang, Z.~Chen, and T.~Yoshioka, ``Minimum bayes risk training for end-to-end speaker-attributed asr,'' in \emph{ICASSP 2021 - 2021 IEEE International Conference on Acoustics, Speech and Signal Processing (ICASSP)}, 2021, pp. 6503--6507.

\bibitem{tian-etal-2023-bayes}
\BIBentryALTinterwordspacing
J.~Tian, B.~Yan, J.~Yu, C.~Weng, D.~Yu, and S.~Watanabe, ``Bayes risk ctc: Controllable ctc alignment in sequence-to-sequence tasks,'' in \emph{The Eleventh International Conference on Learning Representations}, 2023.
\BIBentrySTDinterwordspacing

\bibitem{goel-and-byrne-2000-minimum}
\BIBentryALTinterwordspacing
V.~Goel and W.~J. Byrne, ``Minimum bayes-risk automatic speech recognition,'' \emph{Computer Speech \& Language}, vol.~14, no.~2, pp. 115--135, 2000.
\BIBentrySTDinterwordspacing

\bibitem{kumar-byrne-2004-minimum}
\BIBentryALTinterwordspacing
S.~Kumar and W.~Byrne, ``Minimum {B}ayes-risk decoding for statistical machine translation,'' in \emph{Proceedings of the Human Language Technology Conference of the North {A}merican Chapter of the Association for Computational Linguistics: {HLT}-{NAACL} 2004}.\hskip 1em plus 0.5em minus 0.4em\relax Boston, Massachusetts, USA: Association for Computational Linguistics, May 2 - May 7 2004, pp. 169--176.
\BIBentrySTDinterwordspacing

\bibitem{shafran-and-byrne-2004-task}
I.~Shafran and W.~Byrne, ``{Task-specific minimum Bayes-risk decoding using learned edit distance},'' in \emph{{Interspeech 2004}}, 2004, pp. 1945--1948.

\bibitem{xu-etal-2010-improved}
H.~Xu, D.~Povey, L.~Mangu, and J.~Zhu, ``An improved consensus-like method for minimum bayes risk decoding and lattice combination,'' in \emph{2010 IEEE International Conference on Acoustics, Speech and Signal Processing}, 2010, pp. 4938--4941.

\bibitem{eikema-aziz-2020-map}
\BIBentryALTinterwordspacing
B.~Eikema and W.~Aziz, ``Is {MAP} decoding all you need? the inadequacy of the mode in neural machine translation,'' in \emph{Proceedings of the 28th International Conference on Computational Linguistics}, D.~Scott, N.~Bel, and C.~Zong, Eds.\hskip 1em plus 0.5em minus 0.4em\relax Barcelona, Spain (Online): International Committee on Computational Linguistics, Dec. 2020, pp. 4506--4520.
\BIBentrySTDinterwordspacing

\bibitem{panayotov-etal-2015-librispeech}
V.~Panayotov, G.~Chen, D.~Povey, and S.~Khudanpur, ``Librispeech: An asr corpus based on public domain audio books,'' in \emph{2015 IEEE International Conference on Acoustics, Speech and Signal Processing (ICASSP)}, 2015, pp. 5206--5210.

\bibitem{godfrey-etal-1992-switchboard}
J.~J. Godfrey, E.~C. Holliman, and J.~McDaniel, ``Switchboard: telephone speech corpus for research and development,'' in \emph{Proceedings of the 1992 IEEE International Conference on Acoustics, Speech and Signal Processing - Volume 1}, ser. ICASSP'92.\hskip 1em plus 0.5em minus 0.4em\relax USA: IEEE Computer Society, 1992, p. 517–520.

\bibitem{mccowan-etal-2005-ami}
I.~McCowan, J.~Carletta, W.~Kraaij, S.~Ashby, S.~Bourban, M.~Flynn, M.~Guillemot, T.~Hain, J.~Kadlec, and V.~Karaiskos, ``The ami meeting corpus,'' in \emph{International Conference on Methods and Techniques}.\hskip 1em plus 0.5em minus 0.4em\relax {UIA}, 2005.

\bibitem{graves-etal-2006-connectionist}
\BIBentryALTinterwordspacing
A.~Graves, S.~Fern\'{a}ndez, F.~Gomez, and J.~Schmidhuber, ``Connectionist temporal classification: labelling unsegmented sequence data with recurrent neural networks,'' in \emph{Proceedings of the 23rd International Conference on Machine Learning}, ser. ICML '06.\hskip 1em plus 0.5em minus 0.4em\relax New York, NY, USA: Association for Computing Machinery, 2006, p. 369–376.
\BIBentrySTDinterwordspacing

\bibitem{koehn-knowles-2017-six}
\BIBentryALTinterwordspacing
P.~Koehn and R.~Knowles, ``Six challenges for neural machine translation,'' in \emph{Proceedings of the First Workshop on Neural Machine Translation}, T.~Luong, A.~Birch, G.~Neubig, and A.~Finch, Eds.\hskip 1em plus 0.5em minus 0.4em\relax Vancouver: Association for Computational Linguistics, Aug. 2017, pp. 28--39.
\BIBentrySTDinterwordspacing

\bibitem{ott-etal-2018-analyzing}
M.~Ott, M.~Auli, D.~Grangier, and M.~Ranzato, ``Analyzing uncertainty in neural machine translation,'' in \emph{International Conference on Machine Learning}, 2018.

\bibitem{eikema-aziz-2022-sampling}
\BIBentryALTinterwordspacing
B.~Eikema and W.~Aziz, ``Sampling-based approximations to minimum {B}ayes risk decoding for neural machine translation,'' in \emph{Proceedings of the 2022 Conference on Empirical Methods in Natural Language Processing}, Y.~Goldberg, Z.~Kozareva, and Y.~Zhang, Eds.\hskip 1em plus 0.5em minus 0.4em\relax Abu Dhabi, United Arab Emirates: Association for Computational Linguistics, Dec. 2022, pp. 10\,978--10\,993.
\BIBentrySTDinterwordspacing

\bibitem{freitag-etal-2023-epsilon}
\BIBentryALTinterwordspacing
M.~Freitag, B.~Ghorbani, and P.~Fernandes, ``Epsilon sampling rocks: Investigating sampling strategies for minimum {B}ayes risk decoding for machine translation,'' in \emph{Findings of the Association for Computational Linguistics: EMNLP 2023}, H.~Bouamor, J.~Pino, and K.~Bali, Eds.\hskip 1em plus 0.5em minus 0.4em\relax Singapore: Association for Computational Linguistics, Dec. 2023, pp. 9198--9209.
\BIBentrySTDinterwordspacing

\bibitem{gumbel0-1954-statistical}
\BIBentryALTinterwordspacing
E.~Gumbel, \emph{Statistical Theory of Extreme Values and Some Practical Applications: A Series of Lectures}, ser. Applied mathematics series.\hskip 1em plus 0.5em minus 0.4em\relax U.S. Government Printing Office, 1954.
\BIBentrySTDinterwordspacing

\bibitem{watanabe-etal-2018-espnet}
S.~Watanabe, T.~Hori, S.~Karita, T.~Hayashi, J.~Nishitoba, Y.~Unno, N.~{Enrique Yalta Soplin}, J.~Heymann, M.~Wiesner, N.~Chen, A.~Renduchintala, and T.~Ochiai, ``{ESPnet: End-to-End Speech Processing Toolkit},'' in \emph{{Interspeech 2018}}, 2018, pp. 2207--2211.

\bibitem{kim-etal-2017-joint-ctc}
S.~Kim, T.~Hori, and S.~Watanabe, ``Joint ctc-attention based end-to-end speech recognition using multi-task learning,'' in \emph{2017 IEEE International Conference on Acoustics, Speech and Signal Processing (ICASSP)}, 2017, pp. 4835--4839.

\bibitem{hori-etal-2017-joint}
\BIBentryALTinterwordspacing
T.~Hori, S.~Watanabe, and J.~Hershey, ``Joint {CTC}/attention decoding for end-to-end speech recognition,'' in \emph{Proceedings of the 55th Annual Meeting of the Association for Computational Linguistics (Volume 1: Long Papers)}, R.~Barzilay and M.-Y. Kan, Eds.\hskip 1em plus 0.5em minus 0.4em\relax Vancouver, Canada: Association for Computational Linguistics, Jul. 2017, pp. 518--529.
\BIBentrySTDinterwordspacing

\bibitem{koehn-2004-statistical}
\BIBentryALTinterwordspacing
P.~Koehn, ``Statistical significance tests for machine translation evaluation,'' in \emph{Proceedings of the 2004 Conference on Empirical Methods in Natural Language Processing}, D.~Lin and D.~Wu, Eds.\hskip 1em plus 0.5em minus 0.4em\relax Barcelona, Spain: Association for Computational Linguistics, Jul. 2004, pp. 388--395.
\BIBentrySTDinterwordspacing

\bibitem{bisani-and-ney-2004-bootstrap}
M.~Bisani and H.~Ney, ``Bootstrap estimates for confidence intervals in asr performance evaluation,'' in \emph{2004 IEEE International Conference on Acoustics, Speech, and Signal Processing}, vol.~1, 2004, pp. 409--412.

\bibitem{deguchi-etal-2024-mbrs}
\BIBentryALTinterwordspacing
H.~Deguchi, Y.~Sakai, H.~Kamigaito, and T.~Watanabe, ``mbrs: A library for minimum {B}ayes risk decoding,'' in \emph{Proceedings of the 2024 Conference on Empirical Methods in Natural Language Processing: System Demonstrations}, D.~I. Hernandez~Farias, T.~Hope, and M.~Li, Eds.\hskip 1em plus 0.5em minus 0.4em\relax Miami, Florida, USA: Association for Computational Linguistics, Nov. 2024, pp. 351--362.
\BIBentrySTDinterwordspacing

\bibitem{ichihara-etal-2025-theoretical}
\BIBentryALTinterwordspacing
Y.~Ichihara, Y.~Jinnai, K.~Ariu, T.~Morimura, and E.~Uchibe, ``Theoretical guarantees for minimum {B}ayes risk decoding,'' in \emph{Proceedings of the 63rd Annual Meeting of the Association for Computational Linguistics (Volume 1: Long Papers)}, W.~Che, J.~Nabende, E.~Shutova, and M.~T. Pilehvar, Eds.\hskip 1em plus 0.5em minus 0.4em\relax Vienna, Austria: Association for Computational Linguistics, Jul. 2025, pp. 16\,262--16\,284.
\BIBentrySTDinterwordspacing

\bibitem{kamigaito-etal-2025-diversity}
\BIBentryALTinterwordspacing
H.~Kamigaito, H.~Deguchi, Y.~Sakai, K.~Hayashi, and T.~Watanabe, ``Diversity explains inference scaling laws: Through a case study of minimum {B}ayes risk decoding,'' in \emph{Proceedings of the 63rd Annual Meeting of the Association for Computational Linguistics (Volume 1: Long Papers)}, W.~Che, J.~Nabende, E.~Shutova, and M.~T. Pilehvar, Eds.\hskip 1em plus 0.5em minus 0.4em\relax Vienna, Austria: Association for Computational Linguistics, Jul. 2025, pp. 29\,060--29\,094.
\BIBentrySTDinterwordspacing

\end{thebibliography}

\end{document}